\documentclass[pdflatex,sn-mathphys-num]{sn-jnl}

\usepackage{graphicx}%
\usepackage{multirow}%
\usepackage{amsmath,amssymb,amsfonts}%
\usepackage{amsthm}%
\usepackage{mathrsfs}%
\usepackage[title]{appendix}%
\usepackage{xcolor}%
\usepackage{textcomp}%
\usepackage{manyfoot}%
\usepackage{booktabs}%
\usepackage{algorithm}%
\usepackage{algorithmicx}%
\usepackage{algpseudocode}%
\usepackage{listings}%
\usepackage{pgfplots}
\pgfplotsset{compat=1.15} 
\usepackage{filecontents}

\usepackage{tikz}
\usetikzlibrary{arrows.meta, positioning, shapes.geometric}

\usepackage{caption}
\theoremstyle{thmstyleone}%
\theoremstyle{thmstyletwo}%

\theoremstyle{thmstylethree}%
\numberwithin{equation}{section}
\begin{document}

\title[A mathematical model of human population reproduction through marriage]{A mathematical model of human population reproduction through marriage}


\author*[1]{\fnm{Hisashi} \sur{Inaba}}\email{inaba57@u-gakugei.ac.jp}

\author[2]{\fnm{Shoko} \sur{Konishi}}\email{moe@humeco.m.u-tokyo.ac.jp}


\affil*[1]{\orgdiv{Department of Education}, \orgname{Tokyo Gakugei University}, \orgaddress{\street{4-1-1 Nukuikita-machi}, \city{Koganei-shi}, \postcode{184-8501}, \state{Tokyo}, \country{Japan}}}

\affil[2]{\orgdiv{Department of Human Ecology}, \orgname{School of International Health, Graduate School of Medicine, The University of Tokyo}, \orgaddress{\street{7-3-1 Hongo}, \city{Bunkyo-ku}, \postcode{113-0033}, \state{Tokyo}, \country{Japan}}}



\abstract{
 We develop a linear one-sex dynamical model of human population reproduction through marriage. In our model, a woman may marry and divorce multiple times; however, only women who are currently married are assumed to bear children.
The iterative marriage process is formulated as a three-state compartmental model, which is described by a system of McKendrick equations with a marital birth rate function that depends on the duration of marriage and the age at marriage. To examine the impact of changing nuptiality on fertility, we derive new formulas for the reproduction indices. 
In particular, the total fertility rate (TFR) is expressed as the product of the total marriage number and the average total marital fertility.
 Using Japanese vital statistics, we show that our model provides a reasonable estimate of the current TFR and its future trajectory. }




\maketitle

Dedicated to the memory of Professor Masaya Yamaguti (1925--1998), who was a great promoter of the bridge between mathematics and the biological and social sciences, and the most influential mentor of Hisashi Inaba.

\section{Introduction}

In modern societies, it is common for individuals to have their children
within stable unions. For the sake of simplicity, we define \emph{marriage} as any form of stable union for reproductive purposes. Thus, human fertility depends mainly on two factors: the frequency of marriage and the fertility schedule within marriage ({\it marital fertility}).
Therefore, it is important to develop a dynamical model of human populations that can take into account marriage dynamics to understand the population-level effects of changes in nuptiality and fertility.
Nevertheless, dynamical models of marital reproduction are still underdeveloped in demography.
One important reason for this is that the full nonlinear age-structured two-sex models in demography are difficult to analyze and therefore allow only limited mathematical results. For details on nonlinear two-sex models, the reader may refer to \cite{Inaba2000} and \cite{Inaba2017}.
However, demographers need a theoretical framework to integrate marriage dynamics into the classical stable population model that provides a basis for interpreting demographic indices in order to understand the rapid fertility changes in developed countries. In fact, for these practical, applied modeling purposes, it is not essential to consider nonlinear pair formation models. 

In this paper, we present a linear dynamical model of human population reproduction through marriage. We aim to establish the relationships between marriage and fertility indices that allow us to understand the impact of changes in marriage trends. Indeed, changes in marriage trends are key to understanding the {\it second demographic transition} (SDT)\footnote{See \cite{Jones2007}, \cite{Les2010}, \cite{Raymo2022} and \cite{vandeKaa1987} for the discussion of the SDT concept.}  in Japan, which began in the mid-1970s.
The Japanese population has historically relied on legally married couples for reproduction, with minimal childbearing among either unmarried couples or singles.  In 2023, nonmarital births accounted for only 2.3 percent of the total number of births. The average age of marriage for women increased from 24.6 in 1970 to 31.8 in 2023. The proportion of women who had never married by age 50 increased from 3.33 percent in 1970 to 17.81 percent in 2020. The total fertility rate (TFR) decreased from 2.13 in 1970 to 1.15 in 2024. However, the average number of births among married women aged 45-49 in 2021 was 1.81\cite{NIPSSR2025}. 
Thus, Japan's SDT is characterized by delayed marriage, an increase in individuals who never marry, and persistently low levels of nonmarital childbearing. Therefore, its long-term below-replacement fertility since 1974 can largely be attributed to changes in marriage.

In our previous work \cite{Inaba1992, Inaba1995}, we developed a dynamical marriage model in which births occurred only within first marriages, as childbirths from remarried women were considered negligible at the time (although no official statistics directly reported the number of births from remarriages). This model successfully captured the effect of changes in nuptiality on fertility in Japan. The model was later extended into a four-compartment framework to evaluate the impact of delayed adulthood on reproduction and the age structure of the population in Italy \cite{Billari2000}.
We further extended the first-marriage reproduction model to incorporate childbearing through remarriage, using age- and duration-dependent transition rates \cite{Inaba1993, Inaba2017}, in response to the gradual increase in divorce and remarriage rates in Japan. However, the model remained theoretical due to insufficient data, and it was not possible to derive simple approximation formulas for the total fertility rate or the basic reproduction number based on quantum and tempo indices for marriage.

Here we refine the latter model under slightly simplified assumptions to make it more analytically tractable.
Now we assume that newborns are produced only by currently married women, and that divorced or bereaved women can remarry, and the forces of remarriage and divorce/bereavement for women are {\it duration-dependent}, but for simplicity, they are independent of chronological age.   The marital fertility rate depends on the duration of marriage and the age at marriage.
The basic model is formulated as a system of McKendrick partial differential equations with appropriate boundary conditions. Despite its complex appearance, it is shown that the birth rate of the population
satisfies the well-known Lotka renewal integral equation.
We then present new formulas for demographic indices. The total fertility rate and the basic reproduction number are expressed as the product of the indices of nuptiality and marital fertility. Such decomposition formulas for the reproduction indices are the most useful tools for understanding the impact of changes in nuptiality on human fertility.
Next, we derive a new approximation formula that can provide the TFR based on some basic demographic measures, and apply the formula to Japanese vital statistics.
Finally we extend the basic model to account for the qualitative difference between first marriages and remarriages.

\section{The basic model}

In the following, we consider a closed female population in which all births occur within marriage.  We divide the population
into three groups: $u$ (never married single women), $v$ (currently married women) and $w$ (previously married single women).
Let $u(t,a)$ be the age-density of the never married women with age $a$ at time $t$, let $v(t,\tau;\zeta)$ be the density of currently married women at time $t$ with marital duration (time spent in the marriage) $\tau$ and  age at marriage $\zeta$ (their chronological age is given by $\tau+\zeta$) and
let $w(t,\tau;\zeta)$ be the density of the previously married single women at time $t$
with duration $\tau$ since their divorce or bereavement, which occurred at age $\zeta$.

 Let $\lambda(a)$ be the {\it force of first marriage} at age
$a$, let $\gamma(\tau)$ be the
{\it force of dissolution} at the marital duration $\tau$, let $\mu(a)$ be the force of mortality at age $a$, $m(\tau;\zeta)$ be the {\it marital fertility rate} at marriage duration $\tau$
for women who married at age $\zeta$, i.e., $m$ counts the number of newborns (both sexes) per unit time, let $\delta(\tau)$ be the {\it force of remarriage} at duration $\tau$ since divorce or bereavement and let
$\kappa$ be the proportion of female newborns.  
For simplicity, we assume that $\gamma$ and $\delta$ are independent of the chronological age and that the force of mortality is not affected by marital status.

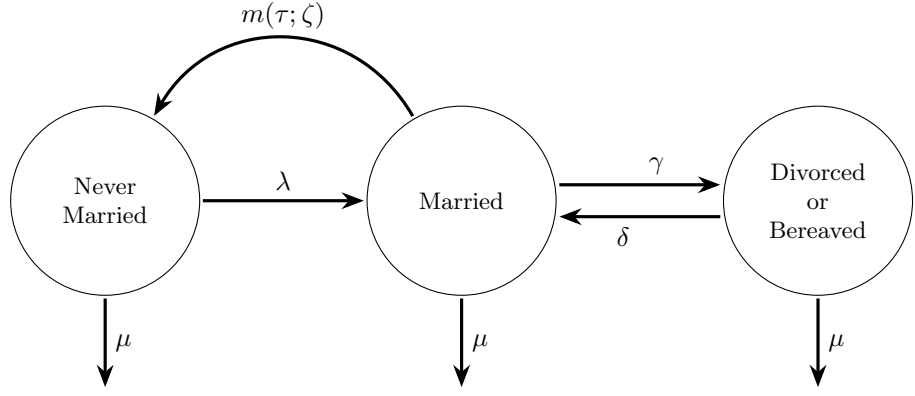
\begin{figure}[htbp]
\centering
\begin{tikzpicture}[
    node distance=2.2cm,
    state/.style={
        circle,
        draw,
        minimum size=2.5cm,
        inner sep=0pt,
        align=center,
        font=\small,
        text width=1.8cm
    },
    arrow/.style={-Stealth,  very thick, shorten >=1pt, shorten <=1pt}, 
    gammaarrow/.style={-Stealth,  very thick, shorten >=1pt, shorten <=1pt}, 
    deatharrow/.style={-Stealth,  very thick, shorten >=1pt, shorten <=1pt} 
]
\node[state] (never) {Never\\Married};
\node[state, right=of never] (married) {Married};
\node[state, right=of married] (divorced) {Divorced\\or\\Bereaved};

\draw[arrow] (never) -- node[above] {$\lambda$} (married);
\draw[gammaarrow] ([yshift=6pt]married.east) -- node[above, pos=0.6] {$\gamma$} ([yshift=6pt]divorced.west);
\draw[arrow] ([yshift=-6pt]divorced.west) -- node[below, pos=0.6] {$\delta$} ([yshift=-6pt]married.east);
\draw[arrow] (married) to[out=120,in=60,looseness=1.2] node[above] {$m(\tau;\zeta)$} (never);

\draw[deatharrow] (never) -- ++(0,-2.5cm) node[midway, right] {$\mu$};
\draw[deatharrow] (married) -- ++(0,-2.5cm) node[midway, right] {$\mu$};
\draw[deatharrow] (divorced) -- ++(0,-2.5cm) node[midway, right] {$\mu$};
\end{tikzpicture}

\
\caption{Three-State Marriage Model: Transitions Between Marital Statuses }
\label{fig:marriage_model_final}
\end{figure}

Then, the dynamical one-sex marriage model is given by:
\begin{equation}\label{basicequ}
\begin{aligned}
&\left(\frac{\partial}{\partial t}+\frac{\partial}{\partial a}\right)
u(t,a)=-[\mu(a)+\lambda(a)]u(t,a),
\cr
&\left(\frac{\partial}{\partial t}+\frac{\partial}{\partial \tau}\right)
v(t,\tau;\zeta)=-[\mu(\tau+\zeta)+\gamma(\tau)]v(t,\tau;\zeta),
\cr
&\left(\frac{\partial}{\partial t}+\frac{\partial}{\partial \tau}\right)
w(t,\tau;\zeta)=-[\mu(\tau+\zeta)+\delta(\tau)]w(t,\tau;\zeta),
\end{aligned}
\end{equation}
whose boundary conditions are given by
\begin{equation}\label{basicbou}
\begin{aligned}
&u(t,0)= \kappa\int_{0}^{\infty}\int_{0}^{\infty}m(\tau;\zeta)
v(t,\tau;\zeta)d\tau d\zeta,
\cr
&v(t,0;\zeta)=\lambda(\zeta)u(t,\zeta)
+\int_{0}^{\zeta}\delta(\tau)w(t,\tau;\zeta-\tau)d \tau,
\cr
&w(t,0;\zeta)=\int_{0}^{\zeta}\gamma(\tau)v(t,\tau;\zeta-\tau)d \tau.
\end{aligned}
\end{equation}
In this system, $u(t,0)$ denotes the number of female newborns per unit time at time $t$; $v(t,0;\zeta)$ represents the number of marriages per unit time and per unit age at time $t$ and age $\zeta$; and $w(t,0;\zeta)$ indicates the number of divorces or bereavements per unit time and per unit age at time $t$ and age $\zeta$.

The basic system \eqref{basicequ} can be integrated
along the characteristic lines, from which it follows that 
\begin{equation}\label{basicsol}
\begin{aligned}
&u(t,a)=\ell(a)\Lambda(a)u(t-a,0),
\cr
&v(t,\tau;\zeta)=
\frac{\ell(\zeta+\tau)}{\ell(\zeta)}\Gamma(\tau)v(t-\tau,0;\zeta),
\cr
&w(t,\tau;\zeta)=
\frac{\ell(\zeta+\tau)}{\ell(\zeta)}\Delta(\tau)w(t-\tau,0;\zeta),
\end{aligned}
\end{equation}
where $\ell$, $\Lambda$, $\Gamma$, $\Delta$ are the survival functions defined by
\begin{equation}\label{survival}
\begin{aligned}
&\ell(a):=e^{-\int_{0}^{a}\mu(\sigma)d\sigma},\quad \Lambda(a):=e^{-\int_{0}^{a}\lambda(\sigma)d\sigma},
\cr
&\Gamma(\tau):=e^{-\int_{0}^{\tau}\gamma(\sigma)d\sigma},\quad \Delta(\tau):=e^{-\int_{0}^{\tau}\delta(\sigma)d\sigma}.
\end{aligned}
\end{equation}

Note that $\ell(a)$ is the life table survival function with the force of mortality $\mu$, and $\Lambda$, $\Gamma$, $\Delta$ are survival functions with hazard rates $\lambda$, $\gamma$ and $\delta$ corresponding to each demographic event. 
In particular, $\Lambda(a)$ denotes the survival function of remaining in the never-married single state at age $a$, $\Gamma(\tau)$ the survival function of remaining in the married state at marital duration $\tau$, and $\Delta(\tau)$ the survival function of remaining in the previously married single state at duration $\tau$ since divorce or bereavement.  

We assume that $\mu$, $\lambda$, $\gamma$ and $\delta$ are nonnegative measurable functions. In addition, we assume that $\int_{0}^{\infty}\mu(a)da=\infty$, which ensures that $\lim_{a \to \infty}\ell(a)=\ell(\infty)=0$. We also assume that $0<\int_{0}^{\infty}\lambda(a)da \le \infty$,  $0<\int_{0}^{\infty}\gamma(\tau)d\tau \le \infty$ and $0<\int_{0}^{\infty}\delta(\tau)d\tau \le \infty$. Thus, the limits $\Lambda(\infty):=\lim_{a \to \infty}\Lambda(a)$, $\Gamma(\infty):=\lim_{a \to \infty}\Gamma(a)$ and $\Delta(\infty):=\lim_{a \to \infty}\Delta(a)$ exist and are each less than one.

\section{Age-specific marriage rate}

It follows from \eqref{basicsol} that the distributions
$u$, $v$ and $w$ can be determined from their boundary values $u(t,0)$, $v(t,0;\zeta)$ and $w(t,0;\zeta)$.

Inserting the expression \eqref{basicsol} into the boundary conditions given in \eqref{basicbou}, we obtain the following system of equations:
\begin{equation}\label{vw1}
\begin{aligned}
&v(t,0;\zeta)=\lambda(\zeta)\Lambda(\zeta)\ell(\zeta)B(t-\zeta)
+\int_{0}^{\zeta}\delta(\tau)\Delta(\tau)\frac{\ell(\zeta)}{\ell(\zeta-\tau)}w(t-\tau,0;\zeta-\tau)d \tau,
\cr
&w(t,0;\zeta)=\int_{0}^{\zeta}\gamma(\tau)\Gamma(\tau)\frac{\ell(\zeta)}{\ell(\zeta-\tau)}v(t-\tau,0;\zeta-\tau)d \tau.
\end{aligned}
\end{equation}

Let $\phi$ and $\psi$ be solutions of the following simultaneous renewal equations:
 \begin{equation}\label{prob}
 \begin{aligned}
&\phi(a)=\lambda(a)\Lambda(a)+\int_{0}^{a}\delta(\tau)\Delta(\tau)\psi(a-\tau)d\tau,
\cr
&\psi(a)=\int_{0}^{a}\gamma(\tau)\Gamma(\tau)\phi(a-\tau)d\tau,
\end{aligned}
\end{equation}
and let $B(t)=u(t,0)$ be the density of newborns at time $t$.
Then it follows that $\ell(\zeta)\phi(\zeta)B(t-\zeta)$ and $\ell(\zeta)\psi(\zeta)B(t-\zeta)$ satisfy \eqref{vw1}.  Therefore, it holds that
\begin{equation}\label{vw2}
\begin{aligned}
&v(t,0;a)=\ell(a)\phi(a)B(t-a),\cr
&w(t,0;a)=\ell(a)\psi(a)B(t-a).
\end{aligned}
\end{equation}
  Thus, we observe that $\phi(a)$ is the incidence rate of marriage at age $a$, and $\psi(a)$ is the incidence rate of divorce or bereavement at age $a$.
We refer to $\phi(a)$ as the {\it age-specific marriage rate} and  $\psi(a)$ as the {\it age-specific dissolution rate}.

 Inserting the second equation of \eqref{prob} into the first equation of \eqref{prob}, we obtain a scalar integral equation for the marriage rate $\phi$:
 \begin{equation}\label{dist}
\phi(a)=\lambda(a)\Lambda(a)+\int_{0}^{a}\theta(\tau)\phi(a-\tau)d\tau,
\end{equation}
where the integral kernel $\theta$ is given by
\begin{equation}
\theta(a):=\int_{0}^{a}\delta(\tau)\Delta(\tau)\gamma(a-\tau)\Gamma(a-\tau)d\tau.
\end{equation}
Once $\phi$ is determined from \eqref{dist}, the dissolution rate $\psi$ is calculated from the second equation in \eqref{prob}.

Let $\hat{f}$ denote the Laplace transform of a function $f$, i.e., 
\begin{equation}
\hat{f}(z):=\int_{0}^{\infty}e^{-za}f(a)da.
\end{equation}  
Taking the Laplace transform of \eqref{dist}, we obtain
\begin{equation}
\hat{\phi}(z):=\hat{\omega}_1(z)+\hat{\omega}_2(z)\hat{\omega}_3(z)\hat{\phi}(z),
\end{equation}
where
\begin{equation}
\omega_1(a):=\lambda(a)\Lambda(a), ~\omega_2(a):=\gamma(a)\Gamma(a), ~\omega_3(a):=\delta(a)\Delta(a),
\end{equation}
Therefore, we have
 \begin{equation} \label{Lap}
 \hat{\phi}(z)=\frac{\hat{\omega}_1(z)}{1-\hat{\omega}_2(z)\hat{\omega}_3(z)}
 \end{equation}
 From \eqref{Lap}, it follows that
 \begin{equation}
 \hat{\phi}(0)=\int_{0}^{\infty}\phi(a)da=\frac{\hat{\omega}_1(0)}{1-\hat{\omega}_2(0)\hat{\omega}_3(0)},
 \end{equation}
where
\begin{equation}
\hat{\omega}_1(0)=1-\Lambda(\infty), ~\hat{\omega}_2(0):=1-\Gamma(\infty), ~\hat{\omega}_3(0):=1-\Delta(\infty).
\end{equation}
Thus $1-\Lambda(\infty)$ is the {\it proportion of ever-married} (PEM), $1-\Gamma(\infty)$ is the {\it proportion of ever-divorced/bereaved} (PED) and $1-\Delta(\infty)$ is the {\it proportion of ever-remarried} (PER).
Therefore, the {\it total marriage number} (TMN) is given by
\begin{equation}\label{TMN}
\int_{0}^{\infty}\phi(a)da=\frac{1-\Lambda(\infty)}{1-(1-\Delta(\infty))(1-\Gamma(\infty))}.
 \end{equation}
The TMN is the expected
total number of marriages per woman in her lifetime under the assumption of no mortality.
From \eqref{prob} and \eqref{TMN}, it follows that
\begin{equation}
\int_{0}^{\infty}\psi(a)da=\frac{(1-\Lambda(\infty))
(1-\Gamma(\infty))}{1-(1-\Gamma(\infty))
(1-\Delta(\infty))},
\end{equation}
which is the {\it total dissolution number} (TDN).
Next, let 
\begin{equation}
\eta(a):=\int_{0}^{a}\delta(\tau)\Delta(\tau)\psi(a-\tau)d\tau,
\end{equation}
denote the age-specific remarriage rate.
Then the {\it total remarriage number} (TRN) is calculated as
\begin{equation}\label{TRN}
\begin{aligned}
\int_{0}^{\infty}\eta(a) da&=(1-\Delta(\infty))\int_{0}^{\infty}\psi(a)da \cr
&=\frac{(1-\Lambda(\infty))(1-\Gamma(\infty))
(1-\Delta(\infty))}{1-(1-\Gamma(\infty))
(1-\Delta(\infty))}.
\end{aligned}
\end{equation}
It follows from \eqref{TMN} and \eqref{TRN} that
\begin{equation}\label{PEDPER}
\frac{{\rm TRN}}{{\rm TMN}}=(1-\Gamma(\infty))(1-\Delta(\infty))={\rm PED} \times {\rm PER},
\end{equation} 
which is a useful relation for estimating ${\rm PED} \times {\rm PER}$ from marriage data, because 
a major obstacle to applying the basic model to real data is that we usually do not have data on $\gamma$ and $\delta$.
It should be emphasized again that these demographic indices are calculated under the assumption of no mortality.

\section{Reproduction Indices}

Once we obtain $\phi$ and $\psi$, it follows from \eqref{basicsol} and \eqref{vw2} that the age or duration density functions $u$, $v$ and $w$ are expressed as follows:
\begin{equation}\label{Bsol}
\begin{aligned}
&u(t,a)=\ell(a)\Lambda(a)B(t-a),\cr
&v(t,\tau;\zeta)=\ell(\zeta+\tau)
\Gamma(\tau)\phi(\zeta)B(t-\tau-\zeta),\cr
&w(t,\tau;\zeta)=\ell(\zeta+\tau)\Delta(\tau)
\psi(\zeta)B(t-\tau-\zeta).\cr
\end{aligned}
\end{equation}

Inserting the expression $v$ in \eqref{Bsol} into the boundary condition of $u(t,0)=B(t)$ in \eqref{basicbou}, it follows that
\begin{equation}
\begin{aligned}
B(t)&= \kappa \int_{0}^{\infty}\int_{0}^{\infty}m(\tau;\zeta)
v(t,\tau;\zeta)d\tau d\zeta,\cr
&=\kappa \int_{0}^{\infty}\int_{0}^{\infty}m(\tau;\zeta)\ell(\tau+\zeta)\Gamma(\tau)\phi(\zeta)B(t-\tau-\zeta)d\tau d\zeta.
\end{aligned}
\end{equation}
By changing the order of integrals, we have
\begin{equation}
B(t)=\kappa \int_{0}^{\infty}\!\!\!\int_{0}^{a}m(a-\zeta;\zeta)
\Gamma(a-\zeta)\phi(\zeta)d\zeta \ell(a)B(t-a)da.
\end{equation}
If we define
\begin{equation}
\begin{aligned}
&K(a):= \kappa \beta(a) \ell(a),\cr
&\beta(a):=\int_{0}^{a}m(a-\zeta;\zeta)\Gamma(a-\zeta)\phi(\zeta)d\zeta,
\end{aligned}
\end{equation}
 $\beta(a)$ denotes the age-specific birth rate and $K(a)$ is the net maternity function.  
Then we arrive at Lotka's integral equation (renewal equation) for the birth rate
\begin{equation}\label{lotka}
B(t)=\int_{0}^{\infty}K(a)B(t-a)da.
\end{equation}
It is well known that the renewal equation \eqref{lotka} admits a unique globally defined, locally integrable solution for $t>0$ if the initial data $B(s)$ for $s <0$ is given and $K \in L^1_+(\mathbb R_+)$ \cite{Inaba2017, Lotka1998}.

From the Lotka renewal integral equation \eqref{lotka}, the {\it basic
reproduction number}, denoted by $R_0$, which is also called the {\it net reproduction rate} in traditional demography\footnote{Although $R_0$ is a dimensionless number, it is traditionally referred to as a ``rate'' in demography.}, is calculated for our 
population system as 
\begin{equation}\label{R0}
R_0=\int_{0}^{\infty}K(a)da=\kappa\int_{0}^{\infty}S(\zeta)
\phi(\zeta)\ell(\zeta)d\zeta,
\end{equation}
where $R_{0}$ is the expected total number of female newborns produced
by a woman over her lifetime, and
\begin{equation}\label{SS}
S(\zeta):= \int_{0}^{\infty}m(\tau;\zeta)\Gamma(\tau)
\frac{\ell(\tau+\zeta)}{\ell(\zeta)}d\tau.
\end{equation}
Since
\begin{equation}\label{mv}
\frac{m(\tau;\zeta)v(t,\tau;\zeta)}{v(t-\tau,0;\zeta)}=m(\tau;\zeta)\Gamma(\tau)
\frac{\ell(\tau+\zeta)}{\ell(\zeta)},
\end{equation}
 $S(\zeta)$ is the expected total number of children produced per marriage at age $\zeta$ (net total marital fertility rate). The formula \eqref{R0} was essentially derived by Itoh and Bando, as shown in equations (13) and (14) of \cite{Itoh1989}.

It is well known that the Euler-Lotka equation
\begin{equation}
\hat{K}(\lambda)=\int_{0}^{\infty}e^{-\lambda a}K(a)da=1,
\end{equation}
has a unique real root, denoted by $r_0$, known as the intrinsic growth rate.
From the Renewal Theorem \cite{Inaba2017}, we know that for $t \to \infty$, there exist $q>0$ (which depends on the initial data) and $\epsilon>0$ such that $B(t)=qe^{r_0 t}(1+O(e^{-\epsilon t}))$.  Then it follows that ${\rm sign}(R_0-1)={\rm sign}(r_0)$, the threshold value for population growth is given by $R_0=1$.

Furthermore, the {\it total fertility rate} (TFR) is given by
\begin{equation}
{\rm TFR}=\int_{0}^{\infty}\beta(a)da=\int_{0}^{\infty}T(\zeta)\phi(\zeta)d\zeta,
\end{equation}
where
\begin{equation}\label{T}
T(\zeta):=\int_{0}^{\infty}m(\tau;\zeta)\Gamma(\tau)d\tau.
\end{equation}
Then, $T(\zeta)$ denotes the total number of children born per marriage to women who marry at age $\zeta$, assuming no mortality. The TFR represents the total number of children a woman would bear over her lifetime, also assuming no mortality. Although the TFR is a dimensionless number, we use the term ``TFR'' in accordance with demographic tradition.
In demography, TFR is the most common quantum measure of fertility, and it also exhibits a threshold property such that ${\rm sign}({\rm TFR}-{\rm PRL})={\rm sign}(r_0)$, where ${\rm PRL}:={\rm TFR}/R_0$ is the population replacement level.  
Although PRL may vary over time, it has remained approximately 2.08 in Japan for decades. It is therefore practical to use TFR as a simple indicator of population growth.

We define the normalized frequency distribution of age at marriage
$\Phi(a)$ as
\begin{equation}
\Phi(a):=\frac{\phi(a)}{\int_{0}^{\infty}\phi(z)dz},
\end{equation}
and the {\it average total marital fertility} (ATMF) as
\begin{equation}\label{atmf}
{\rm ATMF}:=\int_{0}^{\infty}T(\zeta)\Phi(\zeta)d\zeta.
\end{equation}
The ATMF represents the average number of children born per marriage, assuming no mortality. This leads to the following decomposition formula for the total fertility rate (TFR):

\begin{equation}\label{tfr}
{\rm TFR}={\rm TMN} \times {\rm ATMF}=\int_{0}^{\infty}\phi(a)da
\times \int_{0}^{\infty}T(\zeta)\Phi(\zeta)d\zeta.
\end{equation}

The decomposition formula \eqref{tfr} is useful for highlighting the effects of nuptiality and marital fertility on the total fertility rate (see \cite{Inaba1992, Inaba1995}), since TMN and ATMF are scalar measures for nuptiality and marital fertility that are independent of the age distribution of the female population.
From \eqref{TMN} and \eqref{tfr}, we obtain the formula 
\begin{equation}\label{tfr2}
{\rm TFR}=\frac{1-\Lambda(\infty)}{1-(1-\Gamma(\infty))
(1-\Delta(\infty))} \int_{0}^{\infty}T(\zeta)\Phi(\zeta)d\zeta.
\end{equation}

\section{Numerical illustration}

As was pointed out by Inaba \cite{Inaba1995}, the total marital fertility rates by age at marriage given by $S$ and $T$ for Japanese women in the 1980s and for historical populations (Japan and England before World War II) can be well approximated by linear functions.  In fact, this linear pattern is also observed in Mediterranean countries in the 1990s \cite{Billari2000}.
 This linearity makes it possible to derive a useful approximation formula for the TFR based on the marriage model.
 
For illustrative purposes, we have calculated $S(\zeta)$ from the vital statistics of Japan as follows: The number of births by maternal age and marriage duration in 1985-2023 was obtained from the Japanese vital statistics data \cite{SBJ2024a}. The number of births with unknown marriage duration was prorated. 
For the same period, the number of marriages by age and calendar year was calculated using the marital status multistate life tables for women born in 1965-2010 \cite{Ishii2024}. Life tables for the 1945-1964 birth cohorts were assumed to be identical to the life table for the 1965 birth cohort. The number of marriages by age for each birth cohort was calculated in proportion to the number of female births for the whole of Japan in the corresponding year \cite{SBJ2024b}. By combining the data for all the birth cohorts, we obtained the number of marriages by age for each year in 1975-2023. According to the method of Itoh and Bando \cite{Itoh1989}, we calculated the modified number (weighted average) of marriages by age at marriage from age 15 to 49 for each year in 1995-2023, which serves as the denominator to calculate the marital fertility rate by marriage duration and age at marriage.

To calculate the net marital fertility rate at marriage duration $\tau$ and the age at marriage $\zeta$, the number of births by age at marriage $\zeta$ and marriage duration $\tau$ was divided by the corresponding (modified) number of marriages that took place $\tau$ years ago at age at marriage $\zeta$ (see \eqref{mv}). Net marital fertility rates for marriage duration 0, 1, 2, $\cdots$, 14 and 15-19 years by age at marriage $\zeta$ were summed to give $S(\zeta)$.
Figure 2 shows that the linear pattern of $S(\zeta)$ is statistically stable over the observed years (1995--2023).

   \begin{figure}
  \centering
  \includegraphics[width=1.0\textwidth]{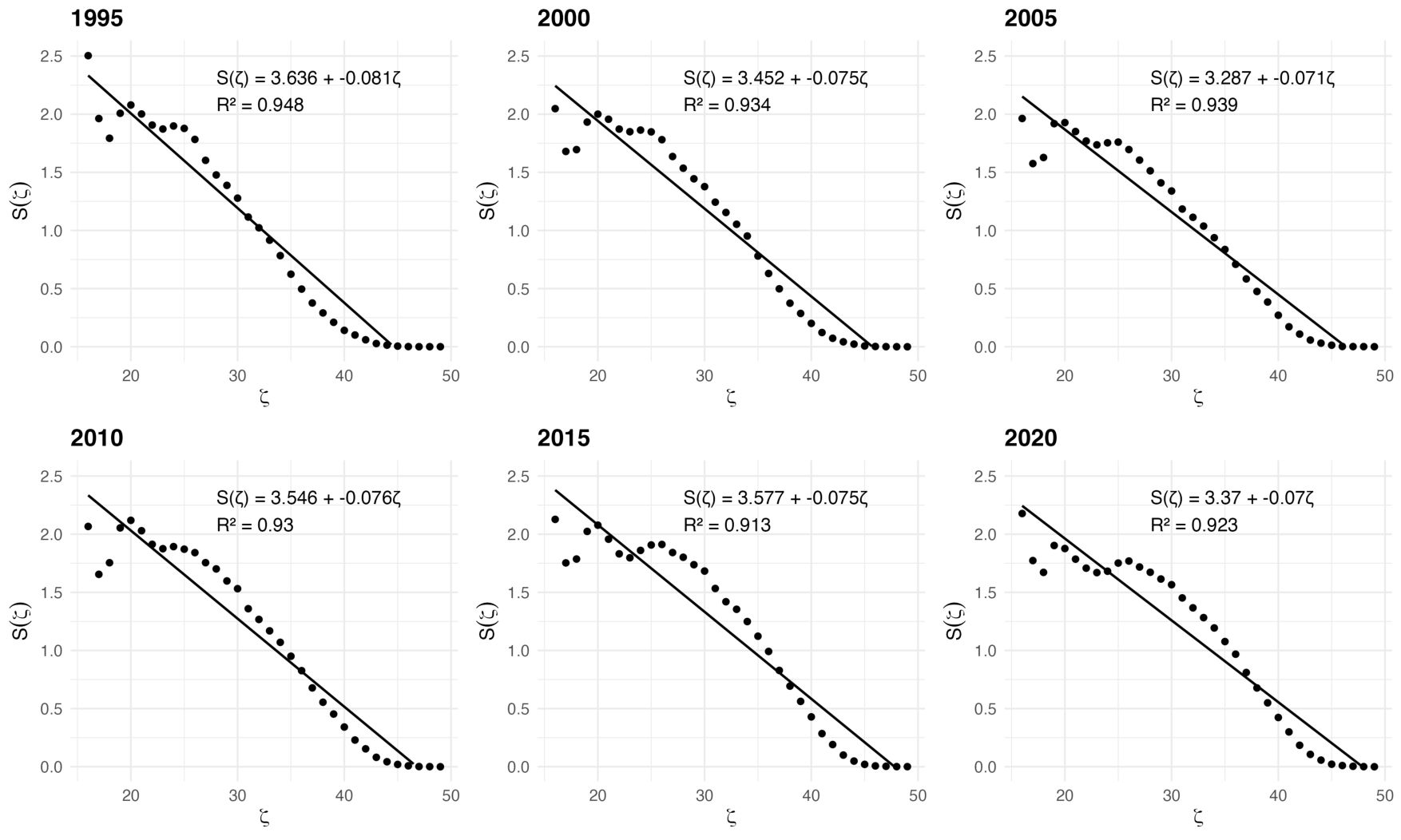} 
     \caption{The net total marital fertility rate by age at marriage $S(\zeta)$ and the regression line from 1995 to 2020.}
  
\end{figure}%

According to the above observations, we can expect $T(\zeta)$ to have an almost linear trend as well.  In fact, if we assume that
\begin{equation}
  S(\zeta)=u+v\zeta+\epsilon_1(\zeta),
\end{equation}
where $\epsilon_1$ is a small residual term, then we have
\begin{equation}\label{A}
\begin{aligned}
T(\zeta)&=S(\zeta)+(T(\zeta)-S(\zeta)) \cr
&=u+v\zeta+\epsilon_2(\zeta),
\end{aligned}
\end{equation}
where
\begin{equation}
\epsilon_2(\zeta):=\epsilon_1(\zeta)+(T(\zeta)-S(\zeta)).
\end{equation}
Then $\epsilon_2(\zeta)$ is very small, because 
\begin{equation}
|T(\zeta)-S(\zeta)| \le T(\zeta) \left|1-\frac{\ell(a_2)}{\ell(a_1)}\right|,
\end{equation}
where $a_1$ denotes the minimum age of marriage and $a_2$ is the maximum reproductive age.  In Japan, for example, the relative difference $|S(\zeta)-T(\zeta)|/T(\zeta)$ is less than one percent, because female mortality during the reproductive period is extremely low.

 Observe that
 \begin{equation}
 \int_{0}^{\infty}(u+v\zeta)\Phi(\zeta)d\zeta=u+va_0,
 \end{equation}
where
\begin{equation}
a_0:=\int_{0}^{\infty}\zeta \Phi(\zeta)d\zeta,
\end{equation}
is the average age of marriage.
Inserting the above expression \eqref{A} into \eqref{tfr2} and neglecting the residual term $\int_{0}^{\infty}\epsilon_2(\zeta)\Phi(\zeta)d\zeta$, we obtain a linear approximation formula
\begin{equation}\label{TFR}
   \begin{aligned}
    {\rm TFR}&    \approx \frac{1-\Lambda(\infty)}{1-(1-\Delta(\infty))(1-\Gamma(\infty))}(u+va_{0})\cr
   & =\frac{{\rm PEM}}{1-{\rm PER}\times {\rm PED}}(u+v a_{0}).
\end{aligned}
\end{equation}

Table 1 shows the TFR calculated using the approximation formula \eqref{TFR}, where we have used the observed proportion of first marriages in total marriages (PFM) as the value of $ 1-{\rm PER}\times {\rm PED}$, because ${\rm PER}\times {\rm PED}$ gives the proportion of remarriages among total marriages in a cohort, if we neglect mortality (see \eqref{PEDPER}).  ${\rm R}^2$ represents the coefficient of determination in the regression of $S(\zeta)$. PEM is calculated as the complement of the proportion of lifetime unmarried observed in each year. pTFR refers to the period TFR observed in the vital statistics for each year.  Table 1 shows that up to 2010 the TFR calculated using \eqref{TFR} differs from the period TFR by only a few percentage points. 
For the 2020 data, it should be noted that people's reproductive patterns were likely affected by the COVID-19 pandemic.

\begin{table}
\centering
\caption{Calculated TFR by the formula \eqref{TFR} for Japan from 1995 to 2020. ${\rm R}^2$ is the coefficient of determination of the linear regression to the data of $S(\zeta)$, PFM is the proportion of first marriages in total marriages and   
pTFR is the period TFR observed from the vital statistics in each year. Source: \cite{NIPSSR2025}}
\setlength{\tabcolsep}{8pt}
\begin{tabular}{|c|c|c|c|c|c|c|c|c|c|}\hline
Year & $u$ & $v$ & ${\rm R}^2$ &$a_0$ & $u+va_0$ & PEM & PFM  & TFR & pTFR \\ \hline
1995 &3.64 &-0.0815 & 0.95& 27.3 & 1.41 & 0.927 &0.884 & 1.48 &1.42\\
2000& 3.45 &-0.0755 & 0.93&28.2 & 1.32 & 0.942 & 0.866& 1.44 &1.36\\
2005 & 3.29 & -0.0701 & 0.94&29.4 & 1.20 & 0.927 & 0.818& 1.36 &1.26\\
2010 & 3.55 & -0.0758 & 0.93&30.3 & 1.25 & 0.893 & 0.815& 1.37 &1.39\\
2015 & 3.58 &-0.0748 & 0.91&31.1 &1.25 & 0.851 & 0.832 & 1.28 &1.45\\
2020 & 3.37 & -0.0704 & 0.92&31.3 & 1.17& 0.822 & 0.832 & 1.15 &1.33\\ \hline
\end{tabular}

\end{table}

In 2023, the National Institute of Population and Social Security Research in Japan published {\it Population Projections for Japan: 2021-2070} \cite{NIPSSR2023}, in which they projected that for Japanese women born in 2005, the total first marriage rate (proportion of ever married, i.e., PEM) will be 0.809, the average age at first marriage will be 28.6, and the cohort TFR for women born in 2005 (the reference cohort) will be 1.29. 
To use the approximation formula \eqref{TFR}, we assume that PEM=0.809, $a_0=31.1$ (observed average age of marriage in 2019), and PER$\times$PED=0.214, corresponding to the fact that the proportion of Japanese marriages in which the wife remarries was 21.4 percent in 2019.  From the 2019 data, we calculated $u=3.53$ and $v=-0.0742$.
Applying \eqref{TFR}, we obtain a TFR of 1.26, which is only 2 percent below the NIPSSR's projected cohort TFR median estimate of 1.29.

As was reported by the National Fertility Survey of Japan \cite{NFS2023}, the completed fertility of Japanese couples with a marriage duration of 15 to 19 years has gradually declined from 2.19 in  1977 to 1.9 in 2021 (a decline of 13 percent), and the framework for reproduction through legal marriage is very robust (the percentage of non-marital births was 2.46 percent in 2023). On the other hand, the period TFR has declined from 2.13 (almost replacement level) in 1970 to 1.2 in 2023, which means that 44 percent of the reproductivity is lost during Japan's second demographic transition \cite{NIPSSR2025}.

\begin{table}
\centering
 \begin{minipage}{\textwidth} 
\centering
\caption{Effects of average age at marriage and PEM on the TFR: $S(\zeta)=3.64-0.0815\zeta$, PFM=0.884}
\setlength{\tabcolsep}{15pt}
\centering
\begin{tabular}{|c|c|c|} \hline
average age at marriage & TFR (PEM=0.927) & TFR (PEM=0.851) \\ \hline
27.3 & 1.48 & 1.36 \\
31.1 & 1.16 & 1.06 \\ \hline
\end{tabular}
\end{minipage}
\end{table}

Under this regime of reproduction, the main reason for the second demographic transition (the decline in the TFR to below the replacement level) in Japan can be explained by the delay in marriage. For example, if we fix the schedule of the net marital fertility rate as $S(\zeta)=3.64-0.0815\zeta$, observed in 1995, and assume that PFM=0.884, we can calculate the TFR using the formula \eqref{TFR} for a given average age at marriage $a_0$ (tempo index of marriage) and PEM (quantum index of marriage), as shown in Table 2, which is based on observed data from 1995 to 2015. The results show that the TFR falls from 1.48 to 1.06, in correspondence with the rise in average age at marriage and the fall in PEM.  However, if only the rise in average age at marriage is considered, the TFR falls to 1.16, which accounts for 76 percent of the decline from 1.48 to 1.06.

\section{Marity progression model}

Finally, we sketch a possible extension of the basic model to make it more realistic.
It is reasonable to assume that human reproductive behavior in the second marriages differ from that in first marriages.  Therefore, we extend the basic model \eqref{basicequ} to account for the difference between the reproductive behavior in the first marriage and that in subsequent marriages.   

We now define {\it marity} as the number of marriages an individual has entered into \cite{Wachter2014}.
Let $v_k(t,\tau;\zeta)$ be the density of currently married women with marity $k$ and marriage duration $\tau$ at time $t$, who entered the $k$-th marriage at age $\zeta$ and
let $w_k(t,\tau;\zeta)$ be the density of the divorced or bereaved single women at time $t$ with marity $k$ and duration $\tau$ since their $k$-th dissolution at age $\zeta$.
Then we can formulate the extended model as follows:

\begin{equation}\label{ext}
\begin{aligned}
&\left(\frac{\partial}{\partial t}+\frac{\partial}{\partial a}\right)
u(t,a)=-[\mu(a)+\lambda(a)]u(t,a),
\cr
&\left(\frac{\partial}{\partial t}+\frac{\partial}{\partial \tau}\right)
v_k(t,\tau;\zeta)=-[\mu(\tau+\zeta)+\gamma_k(\tau)]v_k(t,\tau;\zeta),
\cr
&\left(\frac{\partial}{\partial t}+\frac{\partial}{\partial \tau}\right)
w_k(t,\tau;\zeta)=-[\mu(\tau+\zeta)+\delta_k(\tau)]w_k(t,\tau;\zeta),
\end{aligned}
\end{equation}
\begin{equation}\label{extbou}
\begin{aligned}
&u(t,0)= \kappa \sum_{k=1}^{\infty}\int_{0}^{\infty}\int_{0}^{\infty}m_k(\tau;\zeta)
v_k(t,\tau;\zeta)d\tau d\zeta,
\cr
&v_1(t,0;\zeta)=\lambda(\zeta)u(t,\zeta),\cr
&v_{k+1}(t,0;\zeta)=\int_{0}^{\zeta}\delta_{k}(\tau)w_k(t,\tau;\zeta-\tau)d \tau,
\cr
&w_k(t,0;\zeta)=\int_{0}^{\zeta}\gamma_k(\tau)v_k(t,\tau;\zeta-\tau)d \tau,
\end{aligned}
\end{equation}
where $k=1,2,3,\cdots$ is the marity, $m_k(\tau;\zeta)$ denotes the $k$-th marital fertility, $\delta_k$ the force of the $k$-th remarriage and $\gamma_k$ the force of $k$-th dissolution.

We can define the incidence rates $\phi_{k}(a)$ and $\psi_{k}(a)$ iteratively by
\begin{equation}\label{phi}
\begin{aligned}
&\phi_{1}(a)=\lambda(a)\Lambda(a),\cr
&\phi_{k+1}(a)=\int_{0}^{a}\delta_k(\tau)\Delta_k(\tau)\psi_{k}(a-\tau)d\tau,\cr
&\psi_{k}(a)=\int_{0}^{a}\gamma_k(\tau)\Gamma_k(\tau)\phi_{k}(a-\tau)d\tau,
\end{aligned}
\end{equation}
where $\Delta_k(\tau):=\exp(-\int_{0}^{\tau}\delta_k(x)dx)$ and  $\Gamma_k(\tau):=\exp(-\int_{0}^{\tau}\gamma_k(x)dx)$.
Thus, we have the renewal equation for the birth rate $B$ (the total number of births per unit time) as
\begin{equation}
B(t)=\sum_{k=1}^{\infty}\int_{0}^{\infty}\int_{0}^{a}m_k(\tau;a-\tau)\phi_{k}(a-\tau)\Gamma_k(\tau)d\tau \ell(a)B(t-a)da,
\end{equation}
and the age-specific birth rate
\begin{equation}
\beta(a)=\sum_{k=1}^{\infty}\int_{0}^{a}m_k(\tau;a-\tau)\phi_{k}(a-\tau)\Gamma_k(\tau)d\tau.
\end{equation}
Hence, the total fertility rate is calculated as follows:
\begin{equation}\label{tfrx}
{\rm TFR}=\int_{0}^{\infty}\beta(a)da=\sum_{k=1}^{\infty}\int_{0}^{\infty}\phi_{k}(a)da \int_{0}^{\infty}T_k(\zeta)
\Phi_{k}(\zeta)d\zeta,
\end{equation}
where
\begin{equation}
\begin{aligned}
&\Phi_{k}(a):=\frac{\phi_{k}(a)}{\int_{0}^{\infty}\phi_{k}(\zeta)d\zeta}.\cr
&T_k(\zeta):=\int_{0}^{\infty}m_k(\tau;\zeta)\Gamma_k(\tau)d\tau,
\end{aligned}
\end{equation}
and $T_k(\zeta)$ gives the total number of children produced in the $k$-th marriage with age at marriage $\zeta$.

On the other hand, it follows from \eqref{phi} that
\begin{equation}
\begin{aligned}
&\int_{0}^{\infty}\phi_{1}(a)da=1-\Lambda(\infty),\cr
&\int_{0}^{\infty}\phi_{k+1}(a)da=(1-\Delta_k(\infty))\int_{0}^{\infty}\psi_{k}(a)da,\cr
&\int_{0}^{\infty}\psi_{k}(a)da=(1-\Gamma_k(\infty))\int_{0}^{\infty}\phi_{k}(a)da.
\end{aligned}
\end{equation}
Then we have for $k \ge 1$,
\begin{equation}
\int_{0}^{\infty}\phi_{k+1}(a)da=(1-\Delta_k(\infty))(1-\Gamma_k(\infty))\int_{0}^{\infty}\phi_k(a)da.
\end{equation}

We define the {\it marity progression ratio} at marity $k$, $k=0,1,2,\cdots$ by
\begin{equation}
A_k:=\frac{\int_{0}^{\infty}\phi_{k+1}(a)da}{\int_{0}^{\infty}\phi_{k}(a)da}=(1-\Delta_k(\infty))(1-\Gamma_k(\infty)),
\end{equation}
for $k \ge 1$, and 
\begin{equation}
A_0:=\int_{0}^{\infty}\phi_{1}(a)da=1-\Lambda(\infty).
\end{equation}
Then the marity progression ratio at marity $k$ is the fraction of women who, having reached marity $k$, go on to get married.
Using the marity progression ratio, we have
\begin{equation}
\int_{0}^{\infty}\phi_{k}(a)da=\prod_{j=0}^{k-1}A_{j}.
\end{equation}

Thus, we obtain
\begin{equation}\label{tfrx}
{\rm TFR}=\sum_{k=1}^{\infty}\left(\prod_{j=0}^{k-1}A_j \right)B_k,
\end{equation}
where
\begin{equation}
B_k:=\int_{0}^{\infty}T_k(\zeta)\Phi_{k}(\zeta)d\zeta,
\end{equation}
is the average number of children produced by a woman during her $k$-th marriage.

For practical purposes, it would be sufficient to consider the first and the second marriages to calculate the TFR.  Moreover, in Japan, there are no official statistics on births to remarried women, so it is difficult to estimate $T_k(\zeta)$. For simplicity, if we assume that $T_k(\zeta)=T(\zeta)$ for all marity $k$, it follows that 
\begin{equation}
{\rm TFR} \approx A_0B_1+A_0A_1B_2,
\end{equation}
where
\begin{equation}
 B_k:=\int_{0}^{\infty}T(\zeta)\Phi_{k}(\zeta)d\zeta. 
\end{equation}
If we use the linear approximation $T(\zeta)=u+v\zeta$, we have
\begin{equation}
B_k=u+v a_k,
\end{equation}
where $a_k$ is the average age of the $k$-th marriage.  Then we obtain
\begin{equation}\label{TFR2}
{\rm TFR} \approx A_0 (u+va_1+A_1(u+v a_2)).
\end{equation}

For example, we apply \eqref{TFR2} to the Japanese vital statistics for 2019. From the vital statistics of 2019, we calculated, for all marriages, $u=3.369$ and $v=-0.07$. We also observe that ${\rm PEM}=A_0=0.86$, $a_1=29.6$, $a_2=40.1$ (i.e., the average age at remarriage) and $A_1=0.214$. Then we have ${\rm TFR}=1.22$, which is also close to the median estimate of the future cohort TFR in the NIPSSR projection and also close to the observed TFR of 1.2 in 2023.
In fact, it would be difficult to obtain enough real data for the marity progression model from existing vital statistics, but it is a useful analytical framework for understanding the impact of changes in marital behavior on fertility.

Note that if the reproductive behavior is independent of the marity status, that is, $T_k=T$, $\gamma_k=\gamma$ and $\delta_k=\delta$ for all $k$, we can recover the expression \eqref{tfr2}, that is, 
\begin{equation}
{\rm TFR}=\int_{0}^{\infty}T(\zeta)\phi(\zeta)d\zeta,
\end{equation}
where
\begin{equation}
\phi(\zeta)=\sum_{k=1}^{\infty}\phi_{k}(\zeta),
\end{equation}
and 
\begin{equation}
\begin{aligned}
\int_{0}^{\infty}\phi(\zeta)d\zeta&=(1-\Lambda(\infty))\sum_{k=1}^{\infty}(1-\Delta(\infty))^{k-1}(1-\Gamma(\infty))^{k-1} \cr
&=\frac{1-\Lambda(\infty)}{1-(1-\Delta(\infty))(1-\Gamma(\infty))}.
\end{aligned}
\end{equation}

\section{Concluding remarks}

In this paper, we have extended the marriage model of Inaba \cite{Inaba1992, Inaba1995} so that remarried couples can also have children.  This model is also a revised version of the model in \cite{Inaba1993}.
It should be noted that our model is still a simplification of the complex
dynamics of marriage phenomena in the real world. 

First, as mentioned in the introduction, our
model
neglects the nonlinear interaction between male and female populations. When we consider the bisexual mating process, we need to deal with nonlinear problems \cite{Inaba2000, Inaba2017}.

Second, the force of remarriage $\delta$ and the force of dissolution $\gamma$ are expected to depend on the chronological age in the real world.

Third, our basic model disregards the parity structure of the female population.
In the marity progression model, the same marital fertility rate $m_k(\tau;\zeta)$ applies to all women with marity $k$, so it ignores the heterogeneity of the remarried women with respect to their {\it parity} (the number of live births that a woman has had), even though women's childbearing decisions are strongly influenced by their parity status.
To obtain a marital fertility measure that is independent of the parity distribution, $m_k(\tau;\zeta)$ should be decomposed by the woman's parity status.
It would be an interesting problem to extend our model to include the parity structure, but this is a future challenge.

Finally, our model focuses on reproduction through marriage, which can be widely interpreted as a stable union between the two sexes.  However, reproduction by unmarried individuals or by unstable unions should be considered in more general modeling.
As the mathematical model becomes more complex, it would become more difficult to obtain empirical data that correspond exactly to the model parameters. 
Nevertheless, the model remains a useful tool for exploring the
impact of changes in marital behavior on demographic indicators through
numerical simulations.

\

\bmhead{Acknowledgements}
 We would like to thank the two reviewers for their helpful comments, which significantly improved the manuscript.
We also thank the Princeton University/University of Tokyo Strategic Partnership Teaching and Research Collaboration and Prof. James M. Raymo for making this collaboration possible.

\section*{Declarations}

\begin{itemize}
\item Funding: Hisashi Inaba was supported by JSPS KAKENHI Grant Number 22K03433 and Japan Agency for Medical Research and Development (JP23fk0108685). Shoko Konishi was supported by JSPS KAKENHI Grant Number 22K10484.

\item Conflict of interest/Competing interests: The authors have no relevant financial or non-financial interests to disclose.
\end{itemize}

\end{document}